%
%
\input harvmac

\noblackbox
%

\def\IZ{{\bf Z}}

\def\IR{\relax{\rm I\kern-.18em R}}

\def\ur{U_{\cal  R}(1)}
\def\ar{A^{\cal  R}}

\def\phib{{\bar \phi}}
\def\abstract#1{
\vskip .5in\vfil\centerline
{\bf Abstract}\penalty1000
{{\smallskip\ifx\answ\bigans\leftskip 2pc \rightskip 2pc
\else\leftskip 5pc \rightskip 5pc\fi
\noindent\abstractfont \baselineskip=12pt
{#1} \smallskip}}
\penalty-1000}  
%
%
\def\npb#1#2#3{{\it Nucl. Phys.} {\bf B#1} (#2) #3}
\def\plb#1#2#3{{\it Phys. Lett.} {\bf #1B} (#2) #3}
\def\prl#1#2#3{{\it Phys. Rev. Lett.} {\bf #1} (#2) #3}
\def\prd#1#2#3{{\it Phys. Rev.} {\bf D#1} (#2) #3}

\def\atmp#1#2#3{{\it Adv. Theor. Math. Phys.} {\bf #1} (#2) #3}
\def\cqg#1#2#3{{\it Class. Quant. Grav.} {\bf #1} (#2) #3}
\def\jhep#1#2#3{JHEP {\bf #1} (#2) #3}
\def\hepth#1{{\tt hep-th/#1}}

\def\cR{{\cal R}}

\def\square#1{\mathop{\mkern0.5\thinmuskip
          \vbox{\hrule\hbox{\vrule\hskip#1\vrule height#1 width 0pt\vrule}
          \hrule}\mkern0.5\thinmuskip}}


\lref\juan{J. M. Maldacena, ``The Large $N$ Limit of Superconformal Field
Theories and Supergravity", \atmp{2}{1998}{231}, \hepth{9711200}.}%
\lref\gkp{S. S. Gubser, I. R. Klebanov and A. M. Polyakov, 
``Gauge Theory Correlators
from Non-critical String Theory", \plb{428}{1998}{105}, \hepth{9802109}.}
\lref\wittenone{E. Witten, ``Anti-de-Sitter Space and Holography",
\atmp{2}{1998}{253}, \hepth{9802150}.}
\lref\thooft{G. 't Hooft, ``A Planar Diagram Model For Strong
Interactions", \npb{72}{1974}{461}.}
\lref\coleman{S. Coleman, ``$1/N$", in the proceedings of the 1979
Erice School on Subnuclear Physics and in ``Aspects of Symmetry",
Cambridge University Press.} 
\lref\ahawit{O. Aharony and E. Witten, ``Anti-de Sitter Space and the
Center of the Gauge Group", \jhep{9811}{1998}{018}, \hepth{9807205}.}
\lref\abks{O. Aharony, M. Berkooz, D. Kutasov and N. Seiberg, ``Linear
Dilatons, NS5-branes and Holography", \jhep{9810}{1998}{004},
\hepth{9808149}.} 
\lref\bg{T. Banks and M. B. Green, ``Nonperturbative Effects in
$AdS_5\times S^5$ String Theory and $d=4$ Yang-Mills",
\jhep{9805}{1998}{002}, \hepth{9804170}.}
\lref\bgkr{M. Bianchi, M. B. Green, S. Kovacs and G. Rossi,
``Instantons in Supersymmetric Yang-Mills and D-instantons in IIB
Superstring Theory", \jhep{9808}{1998}{013}, \hepth{9807033}.}
\lref\dkmv{N. Dorey, V. V. Khoze, M. P. Mattis and S. Vandoren,
``Yang-Mills Instantons in the Large $N$ Limit and the AdS/CFT
Correspondence", \plb{442}{1998}{145}, \hepth{9808157}.}
\lref\dhkmv{N. Dorey, T. J. Hollowood, V. V. Khoze, M. P. Mattis and
S. Vandoren, ``Multiinstantons and Maldacena's Conjecture",
\hepth{9810243}.}
\lref\anskeh{D. Anselmi and A. Kehagias, ``Subleading Corrections and
Central Charges in the AdS/CFT Correspondence", \hepth{9812092}.}
\lref\mitone{D. Z. Freedman, S. D. Mathur, A. Matusis and L. Rastelli,
``Correlation Functions in the $CFT_d/AdS_{d+1}$ Correspondence", 
\hepth{9804058}.}
\lref\mittwo{D. Anselmi, D.Z. Freedman, M.T. Grisaru and A.A. Johansen, 
``Universality of the Operator Product Expansion of SCFT$_4$", 
\plb{394}{1997}{329}, \hepth{9608125}.}
\lref\henski{M. Henningson and K. Skenderis, ``The Holographic Weyl
Anomaly", \jhep{9807}{1998}{023}, \hepth{9806087}.}
\lref\gubser{S. S. Gubser, ``Einstein Manifolds and Conformal Field
Theories", \prd{59}{1999}{025006}, \hepth{9807164}.}
\lref\wittenori{E. Witten, ``Baryons and Branes in Anti-de Sitter
Space", \jhep{1998}{9807}{006}, \hepth{9805112}.}
\lref\ks{S. Kachru and E. Silverstein, ``4d Conformal Theories and
Strings on Orbifolds", \prl{80}{1998}{4855}, \hepth{9802183}.}
\lref\sen{A. Sen, ``F Theory and Orientifolds", \npb{475}{1996}{562},
\hepth{9605150}.}
\lref\bds{T. Banks, M. R. Douglas and N. Seiberg, ``Probing F Theory
with Branes", \plb{387}{1996}{278}, \hepth{9605199}.}
\lref\asty{O. Aharony, J. Sonnenschein, S. Theisen and S. Yankielowicz, 
``Field Theory Questions for String Theory Answers",
\npb{493}{1997}{177}, \hepth{9611222}.}
\lref\dls{M. R. Douglas, D. A. Lowe and J. H. Schwarz, ``Probing F-theory 
with Multiple Branes", \plb{394}{1997}{297}, \hepth{9612062}.}
\lref\fs{A. Fayyazuddin and M. Spalinski, 
``Large $N$ Superconformal Gauge Theories
and Supergravity Orientifolds", \npb{535}{1998}{219}, \hepth{9805096}.}
\lref\afm{O. Aharony, A. Fayyazuddin and J. M. Maldacena, ``The Large
$N$ Limit of
${\cal N}=1,2$ Field Theories from Threebranes in F-theory",
\jhep{9807}{1998}{013}, \hepth{9806159}.}
\lref\agw{L. Alvarez-Gaum\'e and E. Witten, ``Gravitational Anomalies", 
\npb{234}{1984}{269}.}
\lref\afgj{D. Anselmi, D. Z. Freedman, M. T. Grisaru 
and A. A. Johansen, ``Nonperturbative Formulas for Central Functions
of Supersymmetric Gauge Theories", \npb{526}{1998}{543},
\hepth{9708042}.}
\lref\ghm{M. B. Green, J. A. Harvey and G. Moore, ``I-brane Inflow and 
Anomalous Couplings on D-branes", \cqg{14}{1997}{47}, \hepth{9605033}.}
\lref\mss{J. Morales, C. Scrucca and M. Serone,
``Anomalous Couplings for D-branes and O-planes",
\hepth{9812071}.}
\lref\stefanski{B. Stefanski, ``Gravitational couplings of D-branes and 
O-planes", \hepth{9812088}.}
\lref\krvn{H. Kim, L. Romans and P. van Nieuwenhuizen, 
``Mass Spectrum of Chiral Ten-dimensional ${\cal N}=2$ Supergravity 
on $S^5$", \prd{32}{1985}{389}.}
\lref\af{G. Arutyunov and S. Frolov, ``Quadratic Action for type IIB 
Supergravity on $AdS_5\times S^5$", \hepth{9811106}.}
\lref\bachas{C. Bachas, ``Lectures on D-branes", \hepth{9806199}.}
\lref\polchinski{J. Polchinski, ``String Theory", Vol. II, 
Cambridge University Press, 1998.}
\lref\ftheory{C. Vafa, ``Evidence for F Theory",
\npb{469}{1996}{403}, \hepth{9602022}.}

\vskip-2cm
\Title{\vbox{
\rightline{\vbox{\baselineskip12pt\hbox{LMU-TPW-99-02}
                                  \hbox{RU-99-2}
                                  \hbox{TAUP-2540-98}
                                   \hbox{\tt hep-th/9901134}}}}}
{A Note on Anomalies in the AdS/CFT Correspondence}
\vskip 0.2cm
\centerline{Ofer Aharony}
\vskip 0.2cm
\centerline{\it Department of Physics and Astronomy}
\vskip-.2cm
\centerline{\it Rutgers University, Piscataway NJ 08855, USA}
\vskip.4cm
\centerline{Jacek Pawe{\l}czyk$^*$, Stefan Theisen}
\vskip 0.2cm
\centerline{\it Institut f\"ur Theoretische Physik} 
\vskip-.2cm
\centerline{\it Universit\"at M\"unchen, 
Theresienstra\ss e 37, 80333 M\"unchen, Germany}
\vskip.4cm
\centerline{Shimon Yankielowicz}
\vskip 0.2cm
\centerline{\it School of Physics and Astronomy}
\vskip-.2cm
\centerline{\it Beverly and Raymond Sackler Faculty of Exact Sciences}
\vskip-.2cm
\centerline{\it Tel-Aviv University, Ramat Aviv, Tel-Aviv 69978, Israel}
\vskip-0.2cm
\abstract{
We test the AdS/CFT
correspondence in the case of a $d=4$ ${\cal N}=2$ SCFT by comparing chiral
anomalies which are of order $N$ in the 't Hooft large $N$ limit. 
These include corrections of order
$1/N$ to the conformal anomaly, thus testing the correspondence beyond
the extreme large $N$ limit. The field theory anomalies are reproduced
by terms in the 7-brane effective action in the bulk.
}
\vskip.5cm
\noindent\hrule
\vskip.2cm\noindent
$^*$On leave from the Institute of Theoretical Physics, Warsaw University.
\hfill\break

\vfill\eject

\parskip=4pt plus 15pt minus 1pt
\baselineskip=15pt plus 2pt minus 1pt

\bigskip

\def\cn{{\cal N}}
 
\newsec{Introduction and Summary}

Over the past year much evidence has accumulated for the conjecture of
\juan\ relating $d$-dimensional conformal field
theories with compactifications of string theory or M theory including
$AdS_{d+1}$. The simplest example of the correspondence is the duality
between the $d=4$ $\cn=4$ $SU(N)$ SYM theory and type IIB string
theory compactified on $AdS_5\times S^5$. This duality relates the
$SU(N)$ gauge theory with coupling $\tau_{YM} \equiv \theta/2\pi +
4\pi i/g_{YM}^2$ to the type IIB compactification with a string
coupling $\tau_s \equiv \chi/2\pi + i/g_s = \tau_{YM}$ and with $N$
units of 5-form flux on the $S^5$, leading to a radius of curvature $R
\sim (g_s N)^{1/4} l_s \sim N^{1/4} l_p$. The correlation functions of
local operators in the conformal field theory are related
\refs{\gkp,\wittenone} to the response of string theory on AdS to
various boundary conditions.

In field theory, it is well-known that correlation functions have a
$1/N$ expansion based on the double-line representation of their Feynman
diagrams \thooft\ (see \coleman\ for a review). This expansion is
valid in the limit of taking large $N$ and small $g_{YM}$ while
keeping $\lambda \equiv g_{YM}^2 N$ finite. A diagram with genus $g$
comes with a power $N^{2-2g}$ as well as some power of $\lambda$. For
$SU(N)$ theories with adjoint fields only closed orientable surfaces
appear, so the perturbative expansion is a double expansion in powers
of $1/N^2$ and $\lambda$, in which the leading term (corresponding to
the planar diagrams) is of order $N^2$.

In the dual AdS theory we find a similar result in the 't Hooft limit
of large $N$ with finite $\lambda \sim g_s N$. In general, correlation
functions in the $AdS_5\times S^5$ string theory are given by a double
expansion in $g_s$ and $\alpha'/R^2 \sim \lambda^{-1/2}$. In the 't
Hooft limit we can write this instead as an expansion in powers of
$1/N \sim g_s (\alpha'/R^2)^2$ where each coefficient has an expansion
in powers of $\lambda^{-1/2}$; obviously a term of some order in the
$g_s$ expansion will have the same order in the $1/N$ expansion. Since
the type IIB string theory includes only closed oriented strings we
find the same general structure as in the field theory, with each
correlation function having an expansion in powers of $1/N^2$, in which
each coefficient is some function of $\lambda$.

The functions of $\lambda$ have different expansions from the point of
view of the field theory and of the string theory, but both expansions
are supposed (if the AdS/CFT correspondence is correct) to give rise
to the same function of $\lambda$ at each order in $1/N^2$. The fact
that one expansion is in powers of $\lambda$ while the other is in
powers of $1/\sqrt{\lambda}$ means that the AdS/CFT correspondence is
a strong/weak coupling duality in this case. Since we do not know how
to make computations to arbitrary order in $\lambda$, this means that
only correlation functions which do not depend on $\lambda$ can be
compared to test the duality. Many such tests have been done by now
for terms of order $N^2$, which on the AdS side may be computed from
tree-level supergravity. However, it is important to test this
matching also at higher orders in $1/N$ in order to make sure that the
duality of \juan\ indeed holds also for finite $N$ and not just in the
large $N$ limit (some evidence for this was provided in
\refs{\ahawit,\abks}). As far as we know, the only computations
performed up to now of higher order terms (in $1/N$) have involved
instanton corrections \refs{\bg,\bgkr,\dkmv,\dhkmv} (some other
corrections were recently discussed but not explicitly computed in
\anskeh).
              
The simplest correlation functions which do not depend on $\lambda$
are those which correspond to global anomalies. Namely, when we put
the theory in backgrounds corresponding to curved space or to gauge
fields coupling to the global currents, some of the global currents
are no longer conserved due to anomalies, and the coefficients in the
expressions for this are integers so they cannot depend on
$\lambda$. In the $\cn=4$ theory there are two such anomalies, which
were both successfully matched to leading order in $1/N^2$. The
anomaly involving three $SU(4)_R$ currents was discussed in
\refs{\wittenone,\mitone}, 
and the conformal anomaly was discussed in \henski.
In the $SU(N)$ SYM theory both anomalies are
proportional to $N^2-1$, and the order $N^2$ term has been matched,
while it is not known how to derive the order $1$ term on the AdS side
(because of our lack of control over the string loop corrections in
this case; note that in general it is easy to obtain the exact
expressions for the anomalies on the field theory side, but it is
non-trivial to get them on the string theory/M theory side). Since the
comparison of the terms of order $1/N^2$ compared to the leading term
seems to be difficult, we would like to examine situations where there
exist also corrections of order $1/N$. In this paper we compute some
corrections of this order and show that they agree between the field
theory and the string theory, providing further evidence for the
AdS/CFT correspondence at finite $N$.

{}From the string theory point of view it is clear that in order to get
diagrams of order $1/N$ one must have either open string diagrams or
non-orientable diagrams. This is well-known also in the field theory
analysis of the 't Hooft limit, where looking at $SO(N)$ or $USp(2N)$
gauge theories leads to non-orientable diagrams with contributions of
order $1/N$, while adding matter in the fundamental representation
(but not in bi-fundamental representations) leads to diagrams with
boundaries which also have contributions of order $1/N$. On
the AdS side such corrections can occur due to orientifolds which lead
to the inclusion of non-orientable worldsheets, or due to D-branes which
lead to worldsheets with boundaries.

Note that in all cases of orbifolds and orientifolds the comparison of
the anomalies in the leading $N^2$ order is straightforward. For
example, it was shown in \gubser\ that the leading term in the
conformal anomaly on the AdS side is inversely proportional to the
volume of the compact space, so that for a $\IZ_k$ orientifold or
orbifold the leading term is $k$ times the $\cn=4$ result. This
obviously agrees with the field theory analysis of D3-branes on
codimension 6 $\IZ_2$
orientifolds \wittenori\ 
which lead to $SO(2N)$, $SO(2N-1)$ or
$USp(2N)$ gauge theories (with a leading anomaly of order $2N^2$), and
with the analysis of D3-branes on codimension 4 $\IZ_k$ orbifolds \ks\ 
which correspond to $SU(N)^k$ gauge theories (with a leading anomaly
of order $kN^2$).

In particular, the analysis of \refs{\henski,\gubser} shows that the
supergravity computation always leads to the two coefficients
appearing in the
conformal anomaly (usually denoted by $a$ and $c$) being equal to each
other, so that duals with a useful supergravity limit can exist only
for theories for which $a=c$ to leading order in $1/N$. We will show
that when higher order corrections are taken into account this no
longer has to hold, so the constraint $a=c$ is only required at the
leading order in $1/N$ (of course, there could also be cases of the
duality which have no good supergravity approximation, in which case
there is no obvious relation between $a$ and $c$).

The simplest case where an order $1/N$ correction to anomalies exists
is the near-horizon limit of D3-branes on an orientifold 3-plane
analyzed in \wittenori. However, in this case the correction on the
string theory side comes from an $RP^2$ diagram which, to our knowledge,
has not been computed yet. Therefore, we will focus here on the
next simplest case, which is the near-horizon limit of D3-branes on a
$\IZ_2$ orientifold 7-plane (with 8 D7-branes stuck on the orientifold
to ensure conformal invariance). The $\cn=2$ superconformal field
theory corresponding to this case was analyzed in
\refs{\sen,\bds,\asty,\dls} and its string theory dual was analyzed in
\refs{\fs,\afm}. We will see that in this case we can compute some of
the order $1/N$ corrections on the string theory side as well as on
the field theory side, by using the effective D7-brane action (whose
leading terms are of order $1/g_s \propto N$ instead of the $1/g_s^2
\propto N^2$ appearing in the SUGRA action), and we will show that
the string theory and field theory results agree to this order. We
have not been able to compute all of the anomalies to this order, and
in particular we do not know how to directly reproduce the conformal
anomaly, but it is related by supersymmetry to the anomaly terms that
we do compute, so supersymmetry guarantees that the conformal anomaly
also agrees. 
Our results may presumably be generalized also to the
other cases involving D7-branes and orientifolds which were
discussed in \refs{\fs,\afm}, for which it is not known
how to compute the anomalies on the field theory side, so the string
theory computation is a prediction of what these anomalies should be.

In section 2 we describe the model and the anomalies of its $U(1)$ 
$R$-current
from the field theory point of view. In section 3 we analyze how these
anomalies are related to Chern-Simons (CS) terms in the effective action
of the D7-branes. 
The results of sections 2 and 3 are compared in section 4. This requires a 
careful fixing of the normalizations,
which involves a comparison of the two-point correlation functions 
of the $R$-current and the flavor current as computed in the field theory 
and as computed via the AdS/CFT correspondence. Unless stated otherwise, 
we will use the conventions of \polchinski. 

\newsec{Anomalies in the $\cn=2$ Superconformal Field Theory}

The model we are considering is the one constructed in \refs{\asty,\dls},
namely the low-energy theory on the worldvolume of $N$ D3-branes
sitting inside eight D7-branes coincident on an orientifold 
7-plane. This theory
is dual \refs{\fs,\afm} to type IIB string theory on $AdS_5\times X^5$ 
where $X^5\simeq S^5/{\bf Z}_2$;
the local operators in the field theory 
can be thought of as living on Mink(3,1)=$\partial(AdS_5)$. 
The D7-branes (and the orientifold 7-plane) 
are wrapped around an $S^3$ which is the fixed point
locus of the ${\bf Z}_2$ orientifold inside $X^5$, and also fill 
the whole of $AdS_5$. The low-energy theory in the bulk of $AdS_5$
includes the gauged $SU(2)_R\times \ur$
$\cn=4$ supergravity, coupled to $SO(8)\times SU(2)_L$ vector multiplets. 
Further details about these models and their dual string theory description
are in \refs{\fs,\afm}.
Some aspects relevant for our discussion will be reviewed below.
The $\ur$ symmetry whose anomalies we study is the one in the inclusion
\eqn\inclusion{
SO(6)\supset SO(4)\times \ur \simeq SU(2)_R\times SU(2)_L\times \ur
}
of the isometry group of $S^5/{\bf Z}_2$ in the isometry group of $S^5$. 

The field theory we are analyzing is a $d=4$, $\cn=2$ gauge theory with
a $USp(2N)$ gauge group and with the following field content:
\eqn\fields{
\eqalign{
\hbox{vector\ multiplets}\qquad & \bf{N(2N+1)}\qquad\,\,\,{\rm (adjoint)}\cr
\hbox{hypermultiplets}\qquad & 4\,\cdot\,\bf{2N} \qquad\qquad\,\, 
                          {\rm(fundamentals)}\cr 
                             & \bf{N(2N-1)-1}
\quad{\rm(antisymmetric,\, traceless).}\cr
}
}
Its global symmetry is $SU(2)_R\times SU(2)_L\times SO(8)\times\ur$,
where $SU(2)_R\times \ur$ is the R-symmetry (which is part of the
$\cn=2$ superconformal algebra).
Here we concentrate on anomalies 
of the $\ur$ $R$-current ${\cal R}^\mu$. 
The form of the unique anomaly-free
${\cal R}^\mu$ current is specified by the 
${\cal R}$-charges for the vector multiplet fermions 
$\lambda_i,\,i=1,2$ ($Q=+1$), 
the bosons $\phi$ ($Q=+2$) in the vector multiplets,
and the matter (hypermultiplet) fermions $\psi\,,\tilde\psi$ ($Q=-1$):
\eqn\rcurrent{
{\cal R}_\mu={1\over 2}\bar\lambda_i\gamma_\mu\gamma_5\lambda_i
-{1\over2}(\bar\psi\gamma_\mu\gamma_5\psi
+\tilde{\bar\psi}\gamma_\mu\gamma_5\tilde\psi)-2i\phib 
{\buildrel \leftrightarrow\over D}_\mu\phi.
}
The factor $1/2$ in front of the fermion bilinears is due 
to the fact that $\lambda_i,\,\psi$ and $\tilde\psi$ are
Majorana spinors. 

The current \rcurrent\ is anomalous when the theory is coupled to gravity.
The anomaly can be calculated using the general result of \agw:
a single Weyl fermion with 
$U(1)$ charge $Q$ contributes to the anomaly as
\eqn\singlean{
\langle\partial_\mu(\sqrt{g}{\cal R}^\mu)\rangle
={Q\over 384\pi^2}(\tilde R R),
}
where
\eqn\deferr{
(\tilde R R)={1\over 2}\epsilon_{\mu\nu\rho\sigma}
R^{\mu\nu}{}_{\delta\gamma}R^{\rho\sigma\delta\gamma}.
}
For the model we are considering, we find
\eqn\fieldanomaly{
\langle\partial_\mu(\sqrt{g}{\cal R}^\mu)\rangle
={2(1-6N)\over 384\pi^2}(\tilde R R).
}
For ${\cal N}=2$ SUSY theories the one-loop result \fieldanomaly\ is not
renormalized\foot{Note that, in contrast to generic ${\cal N}=1$ theories,
the $R$-current 
which is in the super-multiplet of currents is also the one which 
satisfies the Adler-Bardeen theorem.}.
As required by supersymmetry, which relates $\langle\partial_\mu
(\sqrt{g}{\cal R}^\mu)\rangle$ and the conformal anomaly
$\langle g^{\mu\nu}T_{\mu\nu}\rangle$,
the coefficient is proportional to $a-c=(1-6N)/24$, where
$\langle g^{\mu\nu}T_{\mu\nu}\rangle=-a E_4-c I_4$, and $E_4$ and $I_4$ 
are (proportional to) the Euler density and the square of the 
Weyl tensor, respectively (see e.g. \refs{\henski,\gubser}).
Note that $a$ and $c$ are both of order $N^2$ in the large
$N$ limit (as computed for the $\cn=4$ case in \henski), so this
coefficient involves corrections of order $1/N$ compared to the
leading term in $a$ and $c$.

In addition to the above gravitational contribution to the anomalous
divergence of the $U(1)_{\cal R}$ current, there is also a contribution 
from the $\langle {\cal R}JJ\rangle$ triangle diagram, where $J$ is the 
$SO(8)$ flavor current, if we couple the theory to external $SO(8)$
gauge fields. Since only the ``quark'' hypermultiplets are charged
under $SO(8)$, this gives
\eqn\globalan{
\langle\partial_\mu(\sqrt{g}{\cal R}^\mu)\rangle
={2N\over 16\pi^2}(\tilde F F),
}
where $(\tilde F F)={1\over2}\epsilon_{\mu\nu\rho\sigma}
{\rm tr}(F^{\mu\nu}F^{\rho\sigma})$, $F=F^a t^a$,  
and the trace is taken in the fundamental representation.
We will fix the normalization of the $SO(8)$ generators 
later by comparing the 
field  theory two-point function of the $SO(8)$ flavor current with the 
two-point function computed from string theory using
the AdS/CFT correspondence.  
Combining the two contributions \fieldanomaly\ and \globalan\ gives, 
to 
leading order in $1/N$, 
\eqn\fieldtotal{
\langle\partial_\mu(\sqrt{g}{\cal R}^\mu)\rangle
=-{N\over 32\pi^2}[(\tilde R R)-4 (\tilde F F)].}

\newsec{Anomalies from Five-Dimensional Chern-Simons Terms}
 
We will now show how the R-current 
anomalies can be obtained from the string theory
dual of this theory. Our procedure will be completely analogous to the
one used in \wittenone\ --  the anomalies will be related to
Chern-Simons terms in the five-dimensional effective action. 
Chern-Simons terms can arise both from the
dimensionally reduced $d=10$ type IIB supergravity
and from the 7-brane/orientifold plane system. 
The former gives possible anomalous 
contributions to $\langle\cR^3\rangle$ and 
$\langle\cR {\cal J}^2\rangle$, where ${\cal J}$'s are the $SU(2)$
currents. They are of order $N^2$ in the large $N$ expansion (since
the whole $d=10$ supergravity action is of this order). 
We will not consider them here. 
The second source for CS terms are the D7-branes and the orientifold
plane, which are
both wrapped around an $S^3$. As we shall see these terms are of order $N$.

We now focus on the 
five-dimensional terms which arise from dimensional reduction of the
Chern-Simons terms, which appear in the world-volume action of the
7-branes, on the internal
$S^3$. The CS terms for a general Dp-brane are \ghm\ 
\eqn\Dpcs{
\mu_p \int C\wedge\sqrt{\hat A(4\pi^2\alpha' R)}\,{\rm tr}(e^{2\pi\alpha' F})
\Bigl|_{p+1}, 
}
where $\mu_p$ 
is the charge of a single Dp-brane, 
$C=\sum C_{(n)}$ is the sum over the antisymmetric form RR fields 
present in the theory
and\foot{Note that we distinguish between $t^a$ in eq. \globalan\ and 
$T^a$ here.  
Their relative normalization will be fixed in section 4.} 
$F=F^a T^a$
is the field strength of the gauge fields on the Dp-brane. 
In our case these are the $SO(8)$ gauge fields. 
The trace is in the fundamental representation. 
The orientifold plane also gives rise to CS terms on the brane 
world-volume. They have recently been determined to be \refs{\mss,\stefanski}
\eqn\Opcs{
\mu_p'\int C\wedge\sqrt{\hat L(\pi^2\alpha' R)}
\Bigl|_{p+1}, 
}
where
$\mu_p'$ is the charge of an orientifold $p$-plane. 
We have suppressed possible contributions from the (NS,NS) B-field
and also the contribution from the normal bundle, which 
are of no relevance for our considerations. Up to the required order,
\eqn\arlr{
\eqalign{
\hat A(R)&=1+{1\over (4\pi)^2}{1\over12}{\rm tr}(R\wedge R)\,,\cr
\hat L(R)&=1-{1\over (2\pi)^2}{1\over6}{\rm tr}(R\wedge R)\,.\cr}
}

First, we would like to determine the coefficients $\mu_p$ and $\mu_p'$.
Before we take the near-horizon limit, 
the model we are considering is T-dual, say along the $x^8,x^9$ 
directions, to the type I string \sen, which has one space-filling 
orientifold plane and 32 nine-branes. After T-duality we have 
four orientifold 7-planes and 32 seven-branes. This model is equivalent to 
type IIB on Mink(7,1)$\times T^2/(I_{89}\cdot (-1)^{F_L}\cdot\Omega)$
where $\Omega$ is the world-sheet parity and $I_{89}:\,
(x^8,x^9)\to(-x^8,-x^9)$ (see \sen\ for details), which in turn is 
F theory on a particular 
K3. F theory \ftheory\ has 24 7-branes. In our setting four pairs 
of mutually non-local seven-branes 
combine to four orientifold planes. The remaining 16 D7-branes are
equivalent to the 32 seven-branes in the type I description. 
The seven-branes, which are T-duals of the type I nine-branes,  
thus each have half the charge of a type 
IIB D7-brane, i.e. $\mu_p^I=1/[2(2\pi)^p (\alpha')^{(p+1)/2}]$. 
This can also be shown by carefully tracing the brane 
tension through T-duality: due to the fact that in the type I theory 
there are only unoriented strings, one has (see e.g. \bachas) 
$\mu_9^{\rm I}={1\over\sqrt{2}}\mu_9^{\rm II}$. T-dualizing 
on $T^2$ with radii $R$ gives, by comparison of the brane energy per unit 
non-compact volume, 
$e^{-\phi}\mu_9^{\rm I}(2\pi R)^2=e^{-\phi'}\mu_7^{\rm I}$. 
The T-dualized dilaton $\phi'$ is obtained by requiring invariance of 
Newton's constant, i.e. 
${e^{-2\phi'}\over\alpha'^4}{1\over2}(2\pi R')^2
={e^{-2\phi}\over\alpha'^4}(2\pi R)^2$ (the additional factor 
of ${1\over2}$ is explained in \polchinski, p.150).
With $R'=\alpha'/R$ one finds
$e^{\phi'}={1\over\sqrt{2}}{\alpha'\over R^2}e^{\phi}$. 
This leads again to $\mu_7^{\rm I}={1\over2}\mu_7^{\rm II}$.
Charge neutrality of the type I theory requires
$\mu_p'=-2^{p-4}\mu_p$.

Thus,
with eight D7-branes\foot{The multiplicity comes from the trace 
of the zeroth order term in the expansion of the exponent, which is
${\rm tr}(e^{2\pi\alpha' F})
=8+{1\over2}(2\pi\alpha')^2{\rm tr}(F^2)+\cdots$.}
and one orientifold 7-plane one obtains from \Dpcs\ and \Opcs\ the terms
\eqn\sumterms{
{1\over 512\pi^5\alpha'^2}\int C_{(4)}\wedge{\rm tr}(R\wedge R)
+{1\over 128\pi^5\alpha'^2}\int C_{(4)}\wedge {\rm tr}(F\wedge F).
}
The orientifold 7-plane contributes to the second term only
insofar as it leads to an orthogonal 
rather than a unitary gauge group on the D7-branes.

Next, we need to rescale the 4-form field to agree with the
conventions used in the AdS literature (for instance in \af\ which we
will use in section 4). The ten-dimensional supergravity low-energy
effective bulk action is
\eqn\tendaction{
S={1\over2\kappa_{10}^2}\int d^{10}x\sqrt{-g}
\left(R-{1\over 4\cdot 5!}(g_s F_{(5)})^2\right),
}
where $2\kappa_{10}^2=(2\pi)^7 g_s^2\alpha'^4$. We have only kept the fields 
which are relevant for our discussion. Self-duality of $F_{(5)}$ has 
to be imposed as an additional constraint. The ten-dimensional
Einstein equation 
resulting from this action, after imposing self-duality of $F_{5}$ 
(which, in particular, implies $F_{(5)}^2=0$), is
$R_{MN}={1\over 16\cdot 3!}(F_{(5)}^2)_{MN}$. 
The solution we are interested in involves the ten-dimensional metric
and the self-dual five-form field strength. 
Due to the presence of the seven-branes, the solution is not 
$AdS_5\times S^5$ but rather $AdS_5\times X^5$, which is dual to the field
theory described above. The metric is
\eqn\maldalimit{  
ds^2=\alpha'\sqrt{8\pi g_s N}
\left({du^2\over u^2}+u^2 dx_\parallel^2+d\Omega_5^2\right),}
where $d\Omega_5^2$ is the metric on $X^5\simeq S^5/{\bf Z_2}$.
The five-form field is
$F_{abcde}=32\pi\alpha'^2 N \epsilon_{abcde}$ and 
$F_{mnpqr}=32\pi\alpha'^2 N\epsilon_{mnpqr}$, with components 
along the $X^5$ and the $AdS_5$ directions, respectively. 
$\epsilon_{abcde}$ and $\epsilon_{mnpqr}$ are the volume forms
of $X^5$ and $AdS_5$, rescaled to radius one.   
Note the replacement $N\to 2N$ as compared to the $AdS_5\times S^5$
solution. This is due to the fact that we require 
${1\over\sqrt{2}\kappa}\int_{X^5}F_{(5)}=\sqrt{2\pi}N$ rather than 
${1\over\sqrt{2}\kappa}\int_{S^5}F_{(5)}=\sqrt{2\pi}N$, and 
Vol$(X^5)={1\over2}{\rm Vol}(S^5)$. 
To match with the normalizations used in section 4, where we
use supergravity
results which were derived in a normalization where the overall radius of
the space is one (rather than $\alpha'^{1/2}(8\pi g_s N)^{1/4}$), 
and where $F_{(5)}$ is scaled such that the Einstein equations take the 
form $R_{MN}={1\over6}(F_{(5)}^2)_{MN}$, 
we rescale the four-form potential 
$C_{(4)}\to 32\pi\alpha'^2 N C_{(4)}$. After this rescaling,
eq. \sumterms\ becomes
\eqn\csterms{
{N\over 16\pi^4}\int C_{(4)}\wedge{\rm tr}(R\wedge R)
+{N\over 4\pi^4}\int C_{(4)}\wedge {\rm tr}(F\wedge F)\,.
}

To get the CS terms of the $d=5$ theory, 
we have to integrate this expression 
over $S^3$, the fixed locus of the ${\bf Z}_2$ action, which the 
7-branes are wrapped around.  
Let us recall from \krvn\ that the $U(1)_{\cal R}$ gauge boson $A_m^{\cal R}$
is a linear combination of two Kaluza-Klein modes $B_m$ and $\phi_m$,
one coming from the metric components $g_{ma}$ and the other from the  
$C_{(4)}$ field, 
\eqn\modes{
\eqalign{
g_{ma}&\equiv B_m(x) Y_a(y)\,,\cr
C_{mabc}&\equiv \phi_m(x)
\epsilon_{abc}{}^{de} D_d Y_e(y)\,.
}} 
Here 
$x$ and $y$ are coordinates on $AdS_5$ and on $X^5$, respectively, and 
$Y_a(y)$ is a vector spherical harmonic on $X^5$ which we now 
construct.

The metric on $X^5$ is \afm
\eqn\xmetric{
d\Omega_5^2=d\theta^2+\sin^2\theta d\phi^2+\cos^2\theta d\Omega_3^2,
}
with $0\leq\theta\leq\pi/2$, $\phi$ has period $\pi$ due to the
orientifolding, and
$d\Omega_3^2$ is the metric on $S^3$. 
The seven-branes are located at $\theta=0$. The $k=1$ vector spherical harmonics
on $S^5$ are
\eqn\harmonics{
Y^{[ij]}_a=x^{[i}\partial_a x^{j]},
}
where $(x^1)^2+\dots+(x^6)^2=1$ and $a=1,\dots,5$ labels the 
coordinates on $S^5$. 
They satisfy $\square{7pt} Y_a =-4 Y_a$ with 
$\square{7pt}=g^{mn}\nabla_m\nabla_n$.
The relevant harmonic on $X^5$ 
for the $U(1)_{\cal R}$ gauge field is $Y_a^{[56]}\equiv Y^{U(1)}_a$ 
(${\bf Z}_2$ acts as $x^{5,6} \to -x^{5,6}$). In polar 
coordinates it is $\vec Y^{U(1)}={1\over 2}\sin^2\theta\,\hat\phi$.
This leads to 
$C_{mabc}=\phi_m(x)\cos^4\theta\,\omega_{abc}$, where
$\omega_{abc}$ is the volume form on the unit three-sphere.
Note (for later use)
that $\int_{X^5}\sqrt{g}g^{ab}Y_a Y_b\,d^5y={\pi^3\over 24}$.

{}From \krvn\ it follows that the combination $A_m\equiv B_m-16\phi_m$
is a massless vector field, whereas $V_m\equiv B_m+8\phi_m$ is a
massive field. We will set $V_m$ to zero in the following.  Up to a
normalization, which we will fix in section 4, $A_m$ is the
$U(1)_{\cal R}$ gauge field. Denoting by $\ar_m$ the $U(1)_{\cal R}$
field which couples canonically to the current ${\cal R}^\mu$ of the
previous section, we thus identify $\phi_m=-{1\over
24}A_m\equiv\eta\ar_m$ for some constant $\eta$. Then, at
$\theta=0$
\eqn\ca{
C_{mabc}=\eta\,\ar_m \,\omega_{abc},
}
up to some irrelevant additive terms proportional to $V_m$. 
Integrating \csterms\ over $S^3$ we obtain the $d=5$ CS terms
\eqn\csi{
\eta\,{N\over 8\pi^2}
\int_{AdS_5}[A^{\cal R}\wedge{\rm tr}(R\wedge R)
+4 A^{\cal R}\wedge {\rm tr}(F\wedge F)]. }
Under the $U(1)_{\cal R}$ gauge 
transformation $\ar\to \ar+d\Lambda$, this CS term transforms as
\eqn\sugran{
-\eta\,{N\over 8\pi^2}
\int_{{\rm Mink}(3,1)}\Lambda\,[{\rm tr}(R\wedge R)+4{\rm tr}(F\wedge F)],
}
and this can be related to the field theory anomaly
$\langle\partial_\mu(\sqrt{g}{\cal R}^\mu)\rangle$ as in
\refs{\wittenone,\mitone}.
Note the normalizations $\int{\rm tr}(R\wedge R)=-{1\over 2}
\int(\tilde R R)\sqrt{-g}d^4x$ and $\int{\rm tr}(F\wedge F)
={1\over2}\int(F\tilde F)\sqrt{-g}d^4x$.

In the last section we determine $\eta$ by carefully normalizing the
various fields, enabling us to compare the expressions for the anomalies
which we obtained from the four-dimensional field theory \fieldtotal\ 
and the dual string theory \sugran. 

\newsec{Fixing the Normalizations}

We start by verifying that the relative coefficients of the background
gauge and gravitational contributions to the chiral anomaly coincide
in the two computations. Since the numerical relative coefficients are
the same, this reduces to showing that the generators in the
fundamental of $SO(8)$ which appear in \fieldtotal\ and \sugran\ are
normalized in the same way.  To this end we first compute the
two-point function of the $SO(8)$ flavor current in the
field theory, which is
\eqn\flavour{
J^a_\mu(x)=\sum_{i=1}^{2N}\left( 
-{1\over2}\bar\psi_i\gamma_\mu(1-\gamma_5)t^a\psi_i
+\bar\phi_i{\buildrel \leftrightarrow\over {D_{\mu}}} t^a \phi_i\right)\,.}
One finds at one loop
\eqn\flavourtwo{\langle J_\mu^a(x) J_\nu^b(0)\rangle
=2N{\rm tr}(t^a t^b){1\over(2\pi)^4}(\eta_{\mu\nu}\square{7pt}
-\partial_\mu\partial_\nu){1\over x^4}\,.}
Next we need the kinetic energy of the $SO(8)$ field in the 
string theory. It can be obtained by compactifying the 
Dirac-Born-Infeld action of the seven-branes on the $S^3$ around which they 
are wrapped. One finds
\eqn\DBI{
S=-\mu_7\int d^8 x e^{-\phi}{\rm tr}\sqrt{-\det(G_{ab}+2\pi\alpha' F_{ab})}
=\mu_7(2\pi\alpha')^2\int d^8 x e^{-\phi}\sqrt{-G}\,
{1\over4}\,{\rm tr}F^2+\dots}
Normalizing the $SO(8)$ generators as ${\rm tr}(T^a T^b)=\lambda\delta^{ab}$,
rescaling the (induced) metric $G_{ab}\to\alpha'\sqrt{8\pi g_s N}G_{ab}$
and integrating over the unit $S^3$ of volume $2\pi^2$ gives
\eqn\Ssgugra{
S=\lambda{N\over 16\pi^2}\int d^5 x\sqrt{-G}F^a_{\mu\nu}F^{a\mu\nu}
\equiv{1\over 4 
g^2_{SO(8)}}\int d^5 x\sqrt{-G}F^a_{\mu\nu}F^{a\mu\nu}.}
{}From here we find the $SO(8)$ gauge coupling in the low-energy
five-dimensional effective theory
\eqn\coupling{
g_{SO(8)}^2={1\over\lambda}{4\pi^2\over N}\,.}
We now compute the current-current correlation function using the 
AdS/CFT correspondence. We follow \mitone\ and obtain
\eqn\corrsugra{
\langle J_\mu^a(x) J_\nu^b(0)\rangle=\delta^{ab}{1\over 2\pi^2 
g_{SO(8)}^2}
(\eta_{\mu\nu}\square{7pt}-\partial_\mu\partial_\nu){1\over x^4}\,.}
Comparison with the field theory result \flavourtwo\ gives 
\eqn\comparison{
\delta^{ab}{8\pi^2\over g_{SO(8)}^2}=2N{\rm tr}(t^a t^b),}
which leads to ${\rm tr}(T^a T^b)={\rm tr}(t^a t^b)$. This verifies that the 
relative normalizations are indeed the same.

After having shown the agreement of the relative normalization 
of the field theory and the string theory results, we will now
turn to verify that the overall normalization also agrees. 
To this end we must properly normalize
the $U(1)_{\cal R}$ gauge field in the low-energy five-dimensional
action.
The quadratic action for the massless gauge fields has been  
computed in \af, with the result
\eqn\action{
S={4 (2N)^2\over(2\pi)^5}\int_{AdS_5}\!\!\!\! d^5 x\sqrt{-g}\,
{\pi^3\over 24}\cdot {1\over 3}(-{1\over 4}F(A)^2)
\equiv-{1\over 4 g_{SG}^2}\int_{AdS_5}\!\!\!\!d^5x\sqrt{-g}F(A^{\cal R})^2\,.}
$g_{SG}$ is defined to be the $U(1)$ gauge coupling constant of the $d=5$
low-energy theory, $A$ is the massless vector field defined above, and  
$A^{\cal R}$ is the rescaled gauge field 
which couples to ${\cal R}^\mu$.
In the units used here, the overall prefactor comes from the term 
${1\over 2\kappa^2}$ in front of the $d=5$ supergravity action.
To determine $g_{SG}$ we will follow once more the procedure
of \mitone.  This means that we compute the two-point function
$\langle {\cal R}^\mu {\cal R}^\nu\rangle$ in the four-dimensional 
field theory, which
is proportional to the central charge in the ${\cal R}^\mu(x){\cal R}^\nu(0)$
operator product. We find, to leading order in $1/N$,
\eqn\twopoint{
\langle{\cal R}_\mu(x){\cal R}_\nu(0)\rangle=
-{8N^2\over (2\pi)^4}(\square{7pt}\eta_{\mu\nu}
-\partial_\mu\partial_\nu){1\over x^4}.}
This can either be extracted from \mittwo\ or computed directly. 
This one-loop result is exact, due to the superconformal symmetry. 
Comparison with the AdS computation (using again the results of \mitone)
gives $g_{SG}=\pi/N$. Using \action\ and the definition of $\eta$ 
this leads to $\eta=1/2$. Plugging this into \sugran\
we find exact agreement between \fieldtotal\ and 
\sugran, verifying the AdS/CFT correspondence to this order.

\vskip1.5cm
\noindent{\bf Acknowledgements}

\noindent
This work is supported in part by the US-Israel Binational 
Science Foundation, by GIF - the German-Israeli foundation for Scientific
Research, by the Israel Science Foundation, and by
the EEC under TMR contract ERBFMRX-CT96-0045. The work of O.A. was
supported in part by DOE grant DE-FG02-96ER40559. The work of J.P. was
supported in part by
Polish State Committee for Scientific Research (KBN) under contract
2P 03 B03 715 (1998-2000).
J.P. would like to thank the Alexander-von-Humboldt Foundation 
for financial support and S.T. Tel Aviv University for hospitality
during the initial stages of this work. 
We would like to thank S. Kuzenko, A. Schwimmer, C. Scrucca
and in particular C.-S. Chu and J. Maldacena for useful discussions.

\vfill

\listrefs

\bye